\documentclass[fleqn,10pt]{wlscirep}
\usepackage[utf8]{inputenc}
\usepackage[T1]{fontenc}
\graphicspath{{./figures/}}

\title{Ultrafast electron dynamics in a topological surface state observed in two-dimensional momentum space}

\author[1]{J.~Reimann}
\author[2,3]{K.~Sumida}
\author[2]{M.~Kakoki}
\author[5]{K. A. Kokh}
\author[5]{O. E. Tereshchenko}
\author[2,4]{A.~Kimura}
\author[1,*]{J.~Güdde}
\author[1]{U.~H{\"o}fer}

\affil[1]{Fachbereich Physik und Zentrum f{\"u}r
Materialwissenschaften, Philipps-Universit{\"a}t, D-35032 Marburg,
Germany}
\affil[2]{Graduate School of Science, Hiroshima University, 1-3-1 Kagamiyama, Higashi-Hiroshima 739-8526, Japan}
\affil[3]{Materials Sciences Research Center, Japan Atomic Energy Agency, Sayo, Hyogo 679-5148, Japan}
\affil[4]{International Institute for Sustainability with Knotted Chiral Meta Matter, 1-3-1 Kagamiyama, Higashi-hiroshima 739-8526, Japan}
\affil[5]{Rzhanov Institute of Semiconductor Physics and V.S. Sobolev Institute of Geology and Mineralogy, SB RAS, 630090
Novosibirsk, Russian Federation}

\affil[*]{Jens.Guedde@physik.uni-marburg.de}

\begin{abstract}
  We study ultrafast population dynamics in the topological surface
  state of Sb$_2$Te$_3$ in two-dimensional momentum space with time-
  and angle-resolved two-photon photoemission. Linear polarized
  mid-infrared pump pulses are used to permit a direct optical excitation
  across the Dirac point.  We show that this resonant excitation is
  strongly enhanced within the Dirac cone along three of the six
  $\bar{\Gamma}$-$\bar{M}$ directions and results in a macroscopic
  photocurrent when the plane of incidence is aligned along a
  $\bar{\Gamma}$-$\bar{K}$ direction. Our experimental approach makes
  it possible to disentangle the decay of transiently excited
  population and photocurent by elastic and inelastic electron
  scattering within the full Dirac cone in unprecedented detail. This
  is utilized to show that doping of Sb$_2$Te$_3$ by vanadium atoms
  strongly enhances inelastic electron scattering to lower energies, but
  only scarcely affects elastic scattering around the Dirac cone.
\end{abstract}
\begin{document}

\flushbottom
\maketitle
%
%
\thispagestyle{empty}

\section*{Introduction}

The most remarkable properties of the topological surface state (TSS)
of three-dimensional (3D) topological insulators (TIs)
are its Dirac-like, quasi-relativistic energy
dispersion and its chiral spin texture in momentum
($k$-)space\cite{Xia09natphys, Chen09sci2}.
The latter locks momentum and spin of electrons in the TSS and
suggests that momentum scattering within the Dirac cone is strongly
suppressed with even complete absence of direct backscattering
\cite{Hasan10rmp}.
This implies that surface currents are not only automatically
spin-polarized, but also flow ballistically over large distances.
Together with the robustness of the TSS against nonmagnetic
pertubations due its topological protection, this makes these surface
states very promising for use in ultrafast low-loss electronics and
spintronics.
It turned out, however, that not only the application of 3D TIs in
real devices, but already the confirmation of these unique properties
by transport measurements is typically impeded by the dominant role of
bulk carriers which are induced by intrinsic electron- or
hole-doping\cite{Scanlon12am,Wang13prb} in combination with the rather
small band gap of even the prototype 3D TIs such as the binary
chalcogenides Bi$_2$Se$_3$, Bi$_2$Te$_3$, and Sb$_2$Te$_3$.

Direct evidence for long-lasting ballistic surface currents
has been provided by ultrafast pump-probe experiments that combine
surface current generation and its time-resolved detection by
angle-resolved photoelectron spectroscopy (ARPES).
The most direct approach resolves on a subcycle time scale the
momentum distribution of Dirac fermions close to the Fermi
level $E_F$ of Bi$_2$Te$_3$ as they are accelerated by the carrier
wave of a THz pulse \cite{Reimann18nat}.
Its dynamics have been described by the semi-classical Boltzmann
equation which accounts for inelastic scattering as well as for
elastic momentum scattering.
Both corresponding scattering times have been found to be longer than
1~ps which shows that THz-accelerated Dirac fermions may propagate
coherently over several hundred nanometres before under-going
scattering \cite{Reimann18nat}.
In a preceding work, we have shown that such long momentum scattering
times are also inhere in electrons that are optically excited into
the initially unoccupied upper branch of the Dirac cone in
intrinsically p-doped Sb$_2$Te$_3$ \cite{Kuroda16prl}.
This has been accomplished in a time- and angle-resolved two-photon
photoemission (2PPE) experiment by using linear polarized mid-IR laser
pump pulses. These pulses do not only permit a direct optical excitation between the
occupied lower and initially unoccupied upper branch of the TSS, but
also generate a strong asymmetry along a given direction in $k$-space
parallel to the surface concentrated at an energy of a few hundred meV above
$E_F$\cite{Kuroda16prl}.
In both experiments, the momentum distribution of the electrons in the TSS
has been recorded in one direction of the two-dimensional (2D) momentum
space of the surface and the current has been deduced from its asymmetry
for opposite parallel momenta along the direction of current flow.
The extracted momentum scattering times therefore represent effective
phenomenological times for backscattering (180$^\circ$ scattering),
although multiple scattering processes with smaller scattering angles
are in fact involved, because direct backscattering should be
completely suppressed in the TSS\cite{Russman19jpcs}.

Here, we present the investigation of the ultrafast population
dynamics in the TSS of Sb$_2$Te$_3$ after direct optical excitation by
mid-IR pulses in the full 2D momentum space. We will show that the
excitation is strongly enhanced along three of the six
$\bar{\Gamma}$-$\bar{M}$ directions, which allows to monitor the
subsequent redistribution of the electrons within the whole Dirac cone
in unprecedented detail.
We find that the enhancement significantly differs along these three
directions, if a mirror plane ($\bar{\Gamma}$-$\bar{M}$ direction) of
the sample surface is oriented perpendicular to the plane of light
incidence. This confirms that the excitation by linear polarized
mid-IR pulses can in fact generate a macroscopic photocurrent which
is automatically spin polarized due to the spin texture of the TSS.

The time-resolved observation of the decay and the redistribution of
the initially inhomogeneous TSS population in 2D momentum space
allows to disentangle inelastic decay out of the TSS and
elastic momentum scattering within the TSS in great detail.
We demonstrate this capability for pristine and vanadium doped
Sb$_2$Te$_3$ and show how the vanadium atoms affect these two scattering
processes in a distinctly different way.  The dominating mechanism for
inelastic decay in p-doped samples such as intrinsicially p-doped
stoichiometric Sb$_2$Te$_3$ has been identified by conventional 2PPE
experiments to be electron-hole pair creation in the partially filled
valence band\cite{Reimann14prb,Hajlaoui14natcomm,Niesner14prb}.  The
corresponding population lifetime therefore strongly depends on the
position of $E_F$ and it could be shown that it can be strongly
enhanced by Fermi level tuning through doping
\cite{Sumida17scirep,Neupane15prl}.  Consequently, the emergence of
additional states in the bulk band gap can be expected to introduce
further decay channels and we have in fact shown that doping of
Sb$_2$Te$_3$ by vanadium strongly reduces the population lifetime even
for small concentrations of a few percent.
This has been attributed to V-induced impurity states which have been
spectroscopically identified by scanning tunneling spectroscopy
(STS)~\cite{Sumida19njp,Sumida21pss}.
We show here that the momentum scattering is on the other hand
surprisingly almost unaffected by the presence of vanadium atoms,
which demonstrates the robustness of momentum scattering in the TSS
against defects even for magnetic scattering centers.

\section*{Experimental method}

Details of the optical setup are described in
Refs.~\cite{Kuroda16prl,Sumida19njp}.  Electrons were excited above
the Fermi level $E_F$ by pump laser pulses with a photon energy in the
mid-IR ($h\nu_{\rm pump}=0.33$~eV, 100~fs) or in the visible (VIS)
($h\nu_{\rm pump}=2.52$~eV, 80~fs) and subsequently photoemitted by
time-delayed ultraviolet (UV) probe laser pulses ($h\nu_{\rm
  probe}=5.04$~eV, 100~fs) at a repetition rate of 200~kHz. The
chosen UV photon energy just suppresses direct photoemission of the
occupied states but provides full access of the
unoccupied part of the TSS in 2PPE with high dynamic range.  Pump and
probe pulses were $p$-polarized and focused on the sample into a spot
with a diameter of $\sim 100$~$\mu$m at angles of incidence of both
beams close to 45$^\circ$.  The experiments were carried out in a
$\mu$-metal shielded UHV chamber at a base pressure of $4\times
10^{-11}$~mbar with the samples cooled to 110~K after {\it in-situ}
cleaving by the Scotch tape method. Characterization of the samples,
which were grown by the modified vertical Bridgman method with
a rotating heat field~\cite{Kokh14},
is described in detail in Ref.~\cite{Sumida19njp}. Photoelectrons were collected
by a hemispherical analyzer (Scienta DA30) equipped with deflection
plates in the electron lens which makes it possible to aquire
energy-momentum ($E$-$k_x$) maps with the electron momentum $k_x$
oriented along the orientation of the entrance slit of the hemisphere
for varying momentum $k_y$ perpendicular to $k_x$. In this way, the
ultrast dynamics of the optically excited population in the initially
unoccupied band structure can be sequentially mapped in the full
two-dimensional momentum space of the sample surface without moving
the sample.

\section*{Results and discussion}

\subsection*{Homogenous population of the TSS by visible excitation}

Before turning to the discussion of the dynamics of the momentum
distribution of photocurrents excited by mid-IR pump pulses, we first
discuss data taken with visible pump pulses in order to disentangle
the impact of the pump and probe pulses on the energy and momentum
distribution of the detected final states.
The latter is governed by both the transient population of the
intermediate state excited by the pump pulses and the sequential
photoemission by the probe pulses into the detected final states.
Visible pump pulses have been shown to initially excite electrons far
above the TSS and result in an indirect population of the Dirac cone
\cite{Sobota12prl,Reimann14prb}.
This involves many scattering events which completely homogenize the
electron distribution in momentum space \cite{Kuroda16prl}.
Therefore, any momentum dependence of the specral weight of the 2PPE
data can be in this case related to the photoemission probe process
and can be used to correct the momentum-dependent 2PPE data for the
optical matrix element of the probe process as we have
shown in Ref.~\cite{Gudde21pss}.

\begin{figure*}[bth]
    \begin{center}
    \includegraphics[width=\textwidth]{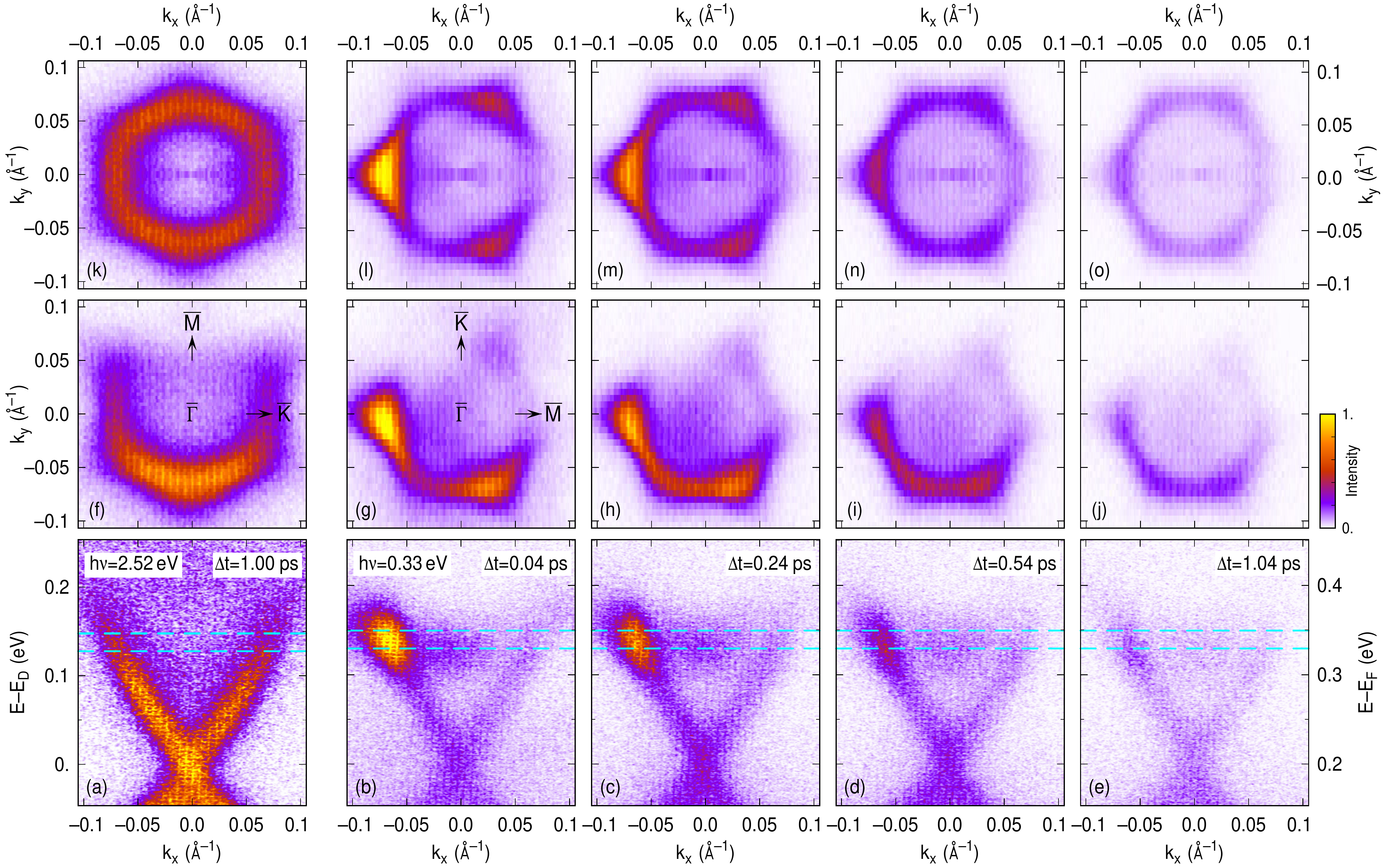}
    \caption{Angle-resolved 2PPE data of Sb$_2$Te$_3$ excited by
      visible ($h\nu=2.52$~eV) and mid-IR ($h\nu=0.33$~eV) pump
      pulses.  a) $E$-$k_x$ map for 2.52-eV excitation and $k_x$ along
      $\bar{\Gamma}$-$\bar{K}$ at a pump-probe delay of $\Delta
      t=1$~ps.  b-e) For 0.33-eV excitation and $k_x$ along
      $\bar{\Gamma}$-$\bar{K}$ at different pump-probe delays as
      indicated. f-j) Corresponding $k_x$-$k_y$ maps integrated over
      energy intervals centered at $E-E_D=140$~meV as is depicted by
      the cyan dashed lines in (a-e). k-o) The same cuts as (f-j) but
      with the intensity corrected for the matrix element of the probe
      transition and symmetrized by mirroring the data at the $k_x$-axis.
      For all data, the plane of light incidence is oriented along the
      $k_y$-axis.
    }\label{fig1}
    \end{center}
\end{figure*}

Figure~\ref{fig1}(a) and (f) show two selected cuts through 2PPE data
of the initially unoccupied part of the TSS taken at $1$~ps after temporal
overlap with visible pump pulses of photon energy $h\nu=2.52$~eV.
At this delay the indirect population of the TSS has reached its
maximum \cite{Reimann14prb}.
The $E$-$k_x$ map for $k_y=0$ depicted in Fig.~\ref{fig1}(a) is
symmetric in $k_x$ and shows the linear dispersion of the TSS along
$\bar{\Gamma}$-$\bar{K}$ with the Dirac point (DP) located 200~meV
above $E_F$.
The TSS is homogeneously populated in energy up to the bulk conduction band
minimum at $E-E_F\approx 330$~meV which feeds the population of the
TSS for several ps \cite{Reimann14prb}.
The $k_x$-$k_y$ map centered at $E-E_D=140$~meV depicted in
Fig.~\ref{fig1}(f) shows the expected warping of the Dirac cone at
this energy with a slight flattening of the linear dispersion along
$\bar{\Gamma}$-$\bar{M}$ \cite{Menshc11jetpl}, but most notably a
strong asymmetry of the 2PPE data with respect to $k_y$ with a
half-moon shaped intensity distribution.
This asymmetry does not result from an inhomogeneous population of the
TSS, but from the oblique incidendence of the p-polarized UV probe
pulses.
This momentum distribution is independent of the sample orientation as
has been tested by azimuthal rotation of the sample and measurements
with two different orientations of the plane-of-incidence.
For $p$-polarized probe pulses incident along the $k_y$-direction as
presented in Fig.~\ref{fig1}, a half moon shaped intensity
distribution indicates that the TSS is dominated by out-of-plane
$sp_z$ orbitals with neglible in-plane contributions which would show
a threefold symmetric pattern \cite{Moser17jelsp}.
The $k_x$-$k_y$ maps can be therefore corrected for the photoemission
probe process by dividing the intensity by $(1-\sin\phi)$ where $\phi$
is the azimuthal angle counting anticlockwise with respect to the
$+k_x$ direction.
This is, however, not applicable for $\phi$ close to 90$^\circ$ and we
apply this correction only for $k_y<0$ and mirror the data with
respect to the $k_x$ axis.
The corrected and symmetrized data depicted in Fig.~\ref{fig1}(k)
shows a homogeneous intensity distribution around the Dirac cone and
in particular no kink at the mirror axis which demonstrates that this
correction accounts well for the matrix element of the UV probe pulses.
This makes it now possible to correct the 2PPE data to reveal the actual population in the
intermediate state also for other pump photon energies, as long as the
same photon energy and polarization of the photemission probe is kept.
Mirroring of the data, however, can be only applied if either the population
is homogeneous in $k$-space or if the plane of probe incidence is oriented
perpendicular to a mirror axis of the sample surface such as the
$\bar{\Gamma}$-$\bar{M}$ direction of the threefold symmetric surface
of Sb$_2$Te$_3$(0001).

\subsection*{Photocurrent generation by mid-IR excitation}

The four rightmost columns of panels in Fig.~\ref{fig1} show 2PPE data
for mid-IR pump pulses which drive a resonant excitation across the
Dirac point \cite{Kuroda16prl} and result in a strongly enhanced
population centered at $E-E_D=140$~meV as can be most clearly seen in
Fig.~\ref{fig1}(b).
Moreover, the mid-IR excitation induces a strong asymmetry with
respect to $k_x$ when the sample is oriented with the
$\bar{\Gamma}$-$\bar{M}$ direction aligned perpendicular to the plane of
probe incidence.
As can be already seen in the raw and uncorrected $k_x$-$k_y$ map
(Fig.~\ref{fig1}(g)), the population is not only enhanced in one
direction, but in three of the six $\bar{\Gamma}$-$\bar{M}$ directions.
Even if the photoemission probe is much less efficient in the
direction of the upper right $\bar{M}$ point, the enhancement is still
faintly visible in the uncorrected data.
The threefold pattern indicates that the optical excitation is
associated with the Sb-Te bonds which have a threefold arrangement in
the unit cell \cite{Glinka15prb}.
The degree of the population enhancement, however, differs in the
three directions.
It is much stronger in direction of the left as compared to that of
the lower right $\bar{M}$ point.
This difference is even further enhanced when the data is corrected
for the photoemission probe as shown in Fig.~\ref{fig1}(l).
This clearly demonstrates that the direct excitation by the mid-IR pulses in
fact generates a macroscopic photocurrent along the $k_x$ direction
while an asymmetry in a $E$-$k_x$ cut as shown Fig.~\ref{fig1}(b) could
also result from a threefold symmetric excitation with equal weight
along the $\bar{\Gamma}$-$\bar{M}$ directions as noticed in
Ref.~\cite{Ketterl18prb}.
However, even in such a case it is possible to gain information about
elastic momentum scattering of the electrons within the Dirac cone by
the investigation of the population balancing along a line in
$k$-space that shows a population
asymmetry\cite{Kuroda16prl,Kuroda17spie}.

\subsection*{Dynamics of momentum scattering}

The time-resolved observation of the redistribution in the full
two-dimensional $k$-space of the surface studied here, however,
provides a much more detailed insight into the initial excitation and
the sequential scattering processes.
This is demonstrated by the data of Fig.~\ref{fig1}(b-e), (g-j) and
(k-o) which shows a time series of cuts through the 2PPE data for
selected delays $\Delta t$ between mid-IR pump and UV probe.
These data show that the population at the resonant excitation energy
of $E-E_D=140$~meV has almost reached its maximum already at $\Delta
t=40$~fs because it is created by a direct optical transition from the
lower to the upper part of the Dirac cone \cite{Kuroda16prl}.
For increasing delay, the whole population at the resonant excitation
energy gradually decreases due to inelastic scattering of the
electrons.
Simultaneously, the population around the Dirac cone becomes more
uniform as can be seen in the $k_x$-$k_y$ maps.
Electrons are elastically scattered in direction within the cone that
were initially less populated.
However, even at $\Delta t = 1.04$~ps, an enhancement of the
population along the $\bar{\Gamma}$-$\bar{M}$ directions is still
clearly visible that shows that elastic scattering, which randomize
the population around the Dirac cone, is weak as compared to inelastic
scattering.
This results from the suppression of momentum scattering due to
the chiral spin structure of the TSS and
is in strong contrast to the electron dynamics in topological
trivial materials such as GaAs where the electrons in the $\Gamma$
valley have been shown to quasiequilibrate in momentum space within 100~fs
\cite{Tanimura21prb}.
The $E$-$k_x$ maps show that the filling of the lower part of the
Dirac cone is delayed with respect to the population at the resonant
excitation energy. This is caused by sequential inelastic scattering
of the directly excited electrons to lower energies.
The intensity at all energies below the resonant excitation energy
still increases from $\Delta t=0.04$~ps to $\Delta t=0.24$~ps and
decays only for longer delays whereas the resonantly populated part at
negative $k_x$ is already reduced at $\Delta t=0.24$~ps.

\begin{figure}[bth]
    \begin{center}
    \includegraphics[width=0.4\textwidth]{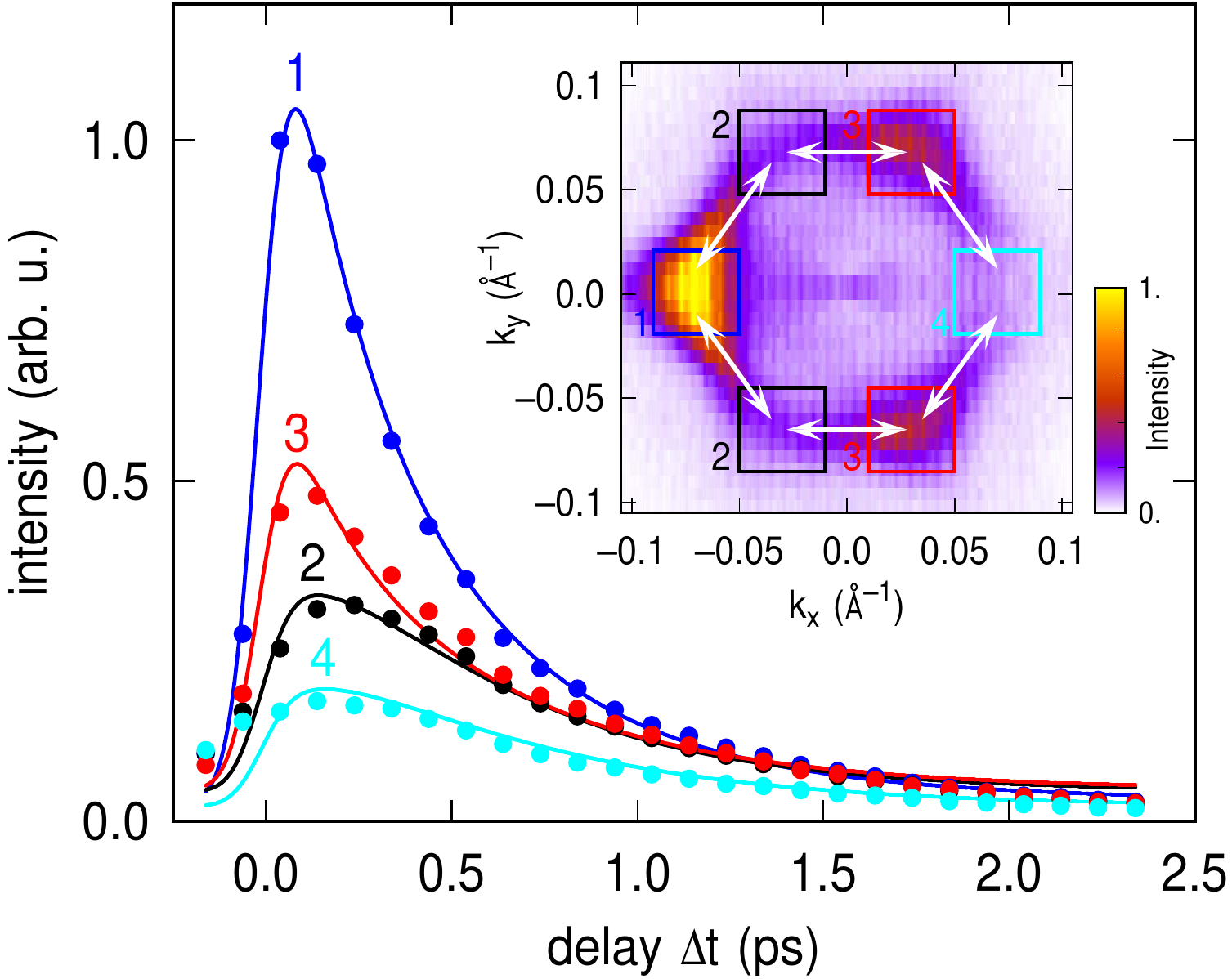}
    \caption{Time-evolution of the 2PPE intensity at different points
      of the $k_x$-$k_y$ maps (c.f. Fig.~\ref{fig1}(l-o)) as is indicated
      by the square integration areas in the inset (data points). The
      solid lines show fits of a rate-equation model for the
      population of the Dirac cone at these points considering
      inelastic electron scattering to lower energies and
      elastic scattering between adjacent areas as indicated by the
      white arrows in the inset.}\label{fig2}
    \end{center}
\end{figure}

The redistribution of the electrons due to elastic scattering within
the TSS is analyzed in more detail in Figure~\ref{fig2}, where we
present the time-evolution of the 2PPE intensity at the resonantly
excited energy along the different $\bar{\Gamma}$-$\bar{M}$
directions.
For this purpose, the 2PPE intensity, which is proportional to the TSS population,
has been integrated over four quadratic regions shown in the inset of Fig.~\ref{fig2}.
The data show that the populations in these regions all decay after their
differently strong initial excitation, but merge with each other at
different delays.
At first, the populations in regions 2 and 3 merge at $\Delta t >
0.5$~ps followed by merging with the initially largest population in region 1 at
$\Delta t > 1$~ps.
The population in region 4 also approaches the others, but even at the
largest observed delay of 2.3~ps it is still the smallest.
The merging shows that the decay of the population in the different
regions is not independent and not only governed by inelastic scattering,
but also accompied by exchange between the regions due to elastic momentum scattering.
The latter can be characterized by a weighted distribution of
scattering angles whereas the spin structure of the TSS suggests an
enhanced probability for small angle scattering and a complete suppression
of direct backscattering \cite{Russman19jpcs}.

We analyze our data by a simplified rate-equation model for the transient
populations $n_i$ in the four regions $i=1\ldots 4$ by considering population loss of all regions with
a common rate $\Gamma^d=1/\tau^d$ and population exchange between neighboring regions
with a common rate $\Gamma^e_{60^\circ}=1/\tau^e_{60^\circ}$ as indicated by the white arrows in the inset of Fig.~\ref{fig2}.
Here, $\tau^d$ and $\tau^e_{60^\circ}$ are the corresponding mean
scattering times for inelastic scattering to lower energies within the TSS or
into the bulk and for $60^\circ$ momentum scattering, respectively.
Considering that the regions 2 and 3 centered at $k_y~-0.06$~\AA$^{-1}$ in the inset of Fig.~\ref{fig2} are identical
to those centered at $k_y~+0.06$~\AA$^{-1}$ due to the mirroring of the data,
the rate equations are given by
\begin{eqnarray}
\frac{dn_1}{dt}&=&A_1{\delta}(t)-{\Gamma^{d}}{n_{1}}+{\Gamma^{e}_{60^\circ}}(2{n_{2}}-2{n_{1}}),\nonumber\\
\frac{dn_2}{dt}&=&A_2{\delta}(t)-{\Gamma^{d}}{n_{2}}+{\Gamma^{e}_{60^\circ}}({n_{1}}+{n_{3}}-2{n_{2}}),\nonumber\\
\frac{dn_3}{dt}&=&A_3{\delta}(t)-{\Gamma^{d}}{n_{3}}+{\Gamma^{e}_{60^\circ}}({n_{2}}+{n_{4}}-2{n_{3}}),\nonumber\\
\frac{dn_4}{dt}&=&A_4{\delta}(t)-{\Gamma^{d}}{n_{4}}+{\Gamma^{e}_{60^\circ}}(2{n_{3}}-2{n_{4}}).
\label{eq:rate_carrier}
\end{eqnarray}
Here, ${\delta}(t)$ is the temporal intensity profile of the Gaussian
shaped mid-IR laser pulse, and the $A_i$ indicate the
different excitation probabilities in the four different regions.
This model represents an extension of the rate-equation model used in
Ref.~\cite{Kuroda16prl} where elastic scattering was characterized by
an effective scattering time for 180$^\circ$ scattering which is in
fact composed of sequential scattering events with smaller scattering
angles. 
The solid lines in Fig.~\ref{fig2} show the best fit
of the data within our model assuming that the 2PPE intensity is
proportional to the TSS population. This fit yields an elastic scattering time of
$\tau^e_{60^\circ}=1.21(15)$~ps for 60$^\circ$ scattering and an
inelastic scattering time of $\tau^d=0.44(10)$ps. The number for
$\tau^e_{60^\circ}$ roughly fits to the previously determined
effective scattering time of 2.5~ps for 180$^\circ$ scattering\cite{Kuroda16prl}.
The latter would correspond to a population exchange only between regions 1 an 4.
The inelastic scattering time is about 30\% smaller as reported in
Ref.~\cite{Kuroda16prl}, but within the variation range observed for
different cleaves of the sample, which has been attributed to variations of
the defect density and of the position of $E_F$ with respect to $E_D$ and the
valence band maximum~\cite{Reimann14prb}.

\subsection*{Impact of V-doping}
\begin{figure*}[tbh]
    \begin{center}
    \includegraphics[width=0.8\textwidth]{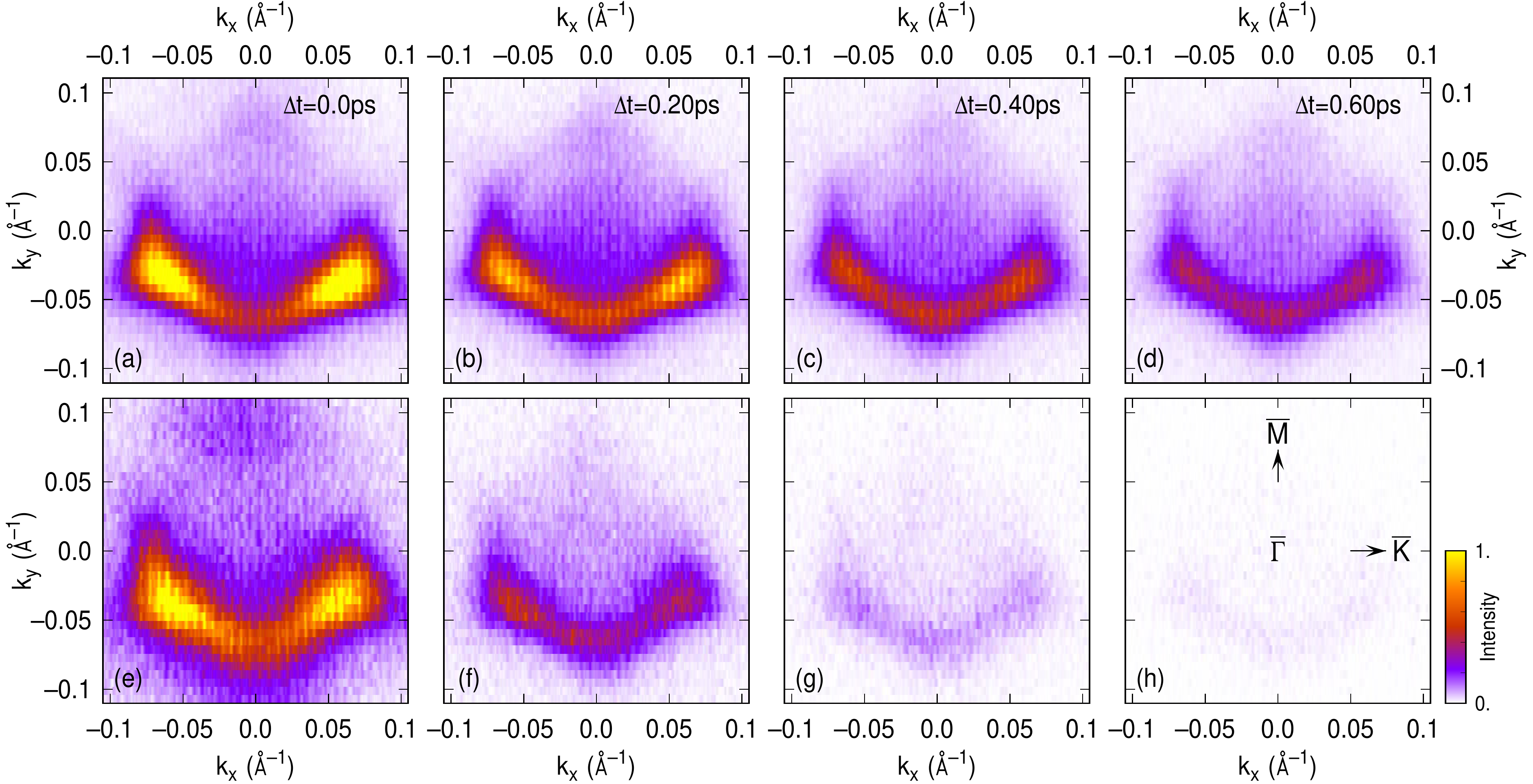}
    \caption{Comparison of the population dynamics between pristine
      Sb$_2$Te$_3$ (a-d) and Sb$_{2-x}V_x$Te$_3$  with
      $x=0.015$ (e-h). The panels show uncorrected
      $k_x$-$k_y$ maps centered at $E-E_D=141$~meV for different time
      delays $\Delta t$. The sample orientation is indicated in panel (h).
      The plane of light incidence is along the
      $k_y$-direction.  }\label{fig3}
    \end{center}
\end{figure*}

In the following, we will show that the capability to observe the
electron dynamics in the full two-dimensional momentum space
makes it possible to disentangle elastic and inelastic electron
scattering, even if no macroscopic photocurrent is generated, as long as
the initial excitation is inhomogeneous in momentum space.
For this purpose, we compare the electron dynamics of
pristine and vanandium doped Sb$_2$Te$_3$ for a different sample orientation.
Recently, we have shown that even small vanadium
concentrations of a few percent drastically reduce the inelastic
scattering time in Sb$_2$Te$_3$, which was attributed to impurity
states\cite{Sumida19njp,Sumida21pss}.
This has been deduced from the observation of the dynamics after
mid-IR excitation in one-dimensional cuts through the TSS band
structure along one surface direction.

\begin{figure}[bth]
    \begin{center}
    \includegraphics[width=0.8\textwidth]{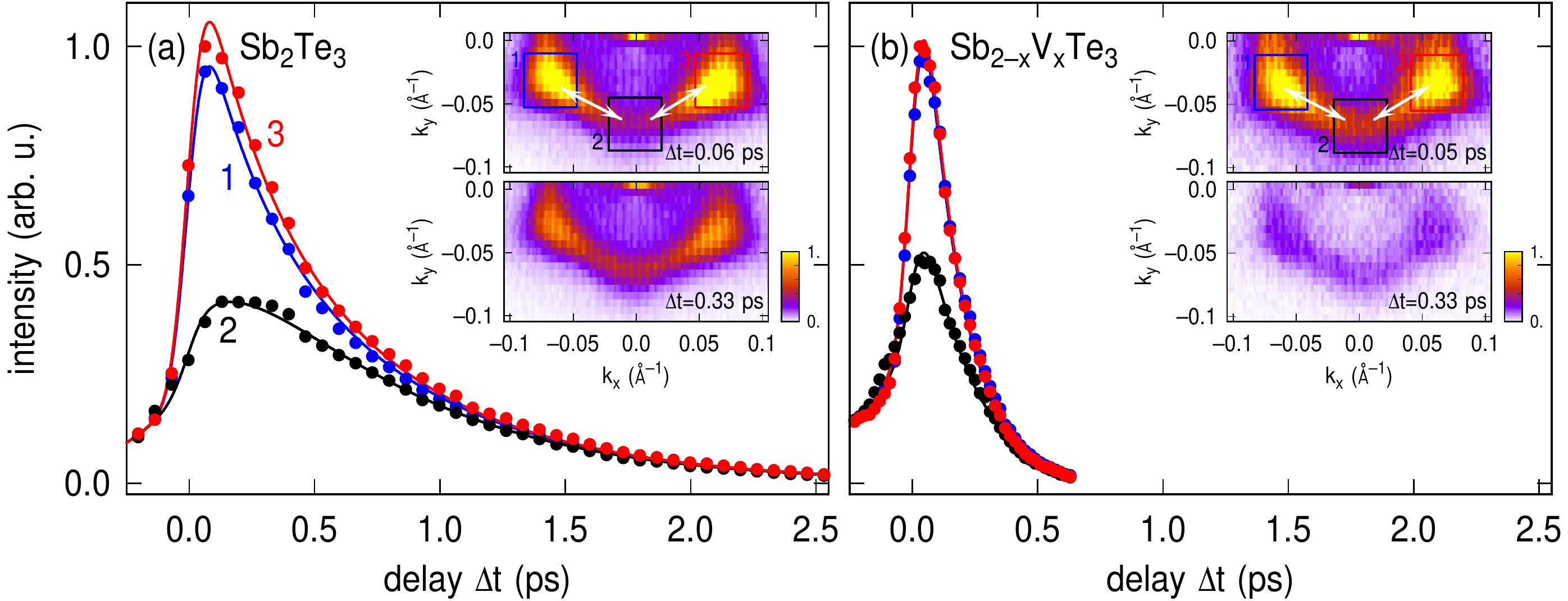}
    \caption{Comparison of the population dynamics between pristine
      (a) Sb$_2$Te$_3$ and (b) Sb$_{2-x}$V$_x$Te$_3$ with $x=0.015$
      with the $\bar{\Gamma}$-$\bar{M}$ direction aligned
      along the plane of incidence.
      The insets show $k_x$-$k_y$ maps centered at
      $E-E_D=140$~meV for two selected time delays $\Delta
      t$. Data points show the intensity integrated over the
      blue and red square areas depicted in the $k_x$-$k_y$ maps.
      Solid lines show fits to the data.
    }\label{fig4}
    \end{center}
\end{figure}

In figure ~\ref{fig3}, we present here a comparison of time-resolved 2PPE data
in the two-dimensional momentum space obtained for pristine
Sb$_2$Te$_3$ and a Sb$_{2-x}$V$_x$Te$_3$ sample with $x=0.015$.
The measurements were done at a temperature of 110~K, which is well
below the Debye temperature of Sb$_2$Te$_3$ ($\Theta_D = 162$~K \cite{Dyck02prb}),
but at the same time well above the Curie temperature of V-doped Sb$_2$Te$_3$ \cite{Sumida19njp}.
Both samples were equally oriented with the $\bar{\Gamma}$-$\bar{M}$
direction aligned along the plane of incidence.
This results in a pure threefold symmetric pattern of the excited
population in the TSS and no macroscopic photocurrent
is generated, as can be most clearly seen in Fig.~\ref{fig3}(a) and
(e) for both samples, respectively.
Information about elastic momentum scattering, however, can still be
extracted from these data because of the initially strong selective enhancement
of the excited population in three of the $\bar{\Gamma}$-$\bar{M}$
whereby only the lower two are clearly observed in the raw data
because of the matrix element of the probe transition.
This is demonstrated in Fig.~\ref{fig4} where we show an analysis 
of the transient TSS population at the resonantly
excited energy for both data after correction of the 2PPE intensity
for the probe matrix element.
Symmetrization by mirroring is not possible for this
threefold symmetric data because the $\bar{\Gamma}$-$\bar{K}$
direction is no mirror plane.
Similar to the analysis shown in Fig.~\ref{fig2}
the 2PPE intensity has been evaluated at the three visible $\bar{M}$-points.
The blue and red solid dots show these data for two of the initially
differently populated $\bar{M}$-points as indicated by the blue and
red rectangles in the $k_x$-$k_y$ map of the inset, respectively. The
solid lines show again fits of the corresponding rate-equation model considering
inelastic decay and 60$^\circ$ elastic momentum scattering.
In order to well describe the initial rise of the transients, we additionally
consider a small contribution of the third image-potential state to the 2PPE intensity,
which is only visible at negative delays (not shown).
It appears at similar final state energies as the TSS but with pump and probe pulses exchanged
\cite{Kuroda17spie,Gudde21pss}. Its rise and decay therefore appears towards negative delays here.

From the data of the pristine sample shown in Fig.\ref{fig4}a, we obtain in this way an
inelastic decay time of $\tau^d=0.68(1)$ps which is very close to the
result of Ref.~\cite{Kuroda16prl}, but longer by $\sim 50\%$ if
compared to the one determined from the data in Fig.~\ref{fig2} due to
the variations for different cleaves as discussed above.  In contrast,
the elastic momentum scattering time $\tau^e_{60^\circ}=1.40(5)$ is
comparable to the one obtained from the data of Fig.~\ref{fig2} which
already indicates that elastic scattering is rather robust against
defects and coupling to the bulk.  Comparing Fig.~\ref{fig4}(a) and
(b) makes it apparent that V-doping strongly reduce the inelastic
scattering time and we determine a four times smaller value of
$\tau^d=0.16(1)$ps as compared to pristine Sb$_2$Te$_3$.  This can be
most probably attributed to the V-induced impurity states that have
been identified by scanning tunneling spectroscopy
(STS)~\cite{Sumida19njp}.  Such impurity states can introduce
additional decay channels for intraband scattering to lower energies
and were shown to persist even in the presence of magnetic order \cite{Sessi16natcom}.
Surprisingly, however, the momentum scattering time is not much
affected by V-doping as can be already seen by comparing the
2D maps of the pristine and V-doped sample in the insets of Fig.~\ref{fig4} taken at 0.33 ps.
Although the overall intensity for the V-doped sample is
more strongly reduced as compared to the pristine sample due
to the stronger inelastic scattering, the inhomogeneity of the
intensity distribution around the Dirac cone is comparable for both samples.
This can be even more clearly seen by comparing the time dependence of the intensity in region 2
with those of region 1 and 3 shown in Fig.~\ref{fig4}b.
It does not still take considerable time for these intensities to merge.
The still long momentum-scattering time is also reflected by the fact
that the maximum difference of these intensities, which is reached around
0.12 ps, is comparable to the one of the pristine sample.
An accelerated momentum-scattering would lead to a faster and stronger
merging of theses intensities.
A fit of the data indeed gives only a slightly reduced momentum-scattering
time of $\tau^e_{60^\circ}=1.0(5)$~ps, but with an enhanced uncertainty due
to the stronger inelastic scattering. 
Such still weak momentum scattering is surprising because it might be expected
that impurities also enhance momentum scattering, in particular if the impurities carry
a magnetic moment.
On the other hand, time-resolved quantum-beat photoelectron spectroscopy
of image-potential states on a Cu(001) surface has shown
that single scattering centers can affect
inelastic decay and momentum scattering in a very different way
depending on the details of the scattering potential \cite{Fauster07pss}.
While CO adatoms mainly lead to the decay of the quantum beats
between different image-potential states due to momentum
scattering \cite{Reuss99prl}, an even much smaller
concentration of Cu adatoms has a stronger impact on the inelastic decay \cite{Fauster00cp}.
Ab initio calculations suggest that for scattering of electrons in a
TSS at single magnetic impurities, the non-magnetic part of the scattering potential
with resonant scattering into defect states dominates
\cite{Russmann18diss}. This is in line with our observation that the
magnetic impurities strongly reduce the inelastic scattering time,
but have only a small impact on elastic momentum scattering.

\section*{Conclusions}
We have experimentally investigated the ultrafast population dynamics
of electrons in the TSS of pristine and V-doped
Sb$_2$Te$_3$ in two-dimensional momentum space after direct optical
excitation by linearly polarized mid-IR pulses using time- and
angle-resolved two-photon photoemission.
We have shown that the population at the resonant excitation energy
in the upper Dirac cone is not uniform in momentum space, but enhanced
along three of the six $\bar{\Gamma}$-$\bar{M}$ directions.
If the the plane of incidence is aligned along one of the
$\bar{\Gamma}$-$\bar{K}$ directions, this enhancement
is not threefold symmetric and corresponds to a macroscopic
photocurrent along a $\bar{\Gamma}$-$\bar{M}$ direction.
The inhomgeneous excitation, together with the detection in the full
two-dimensional momentum space of the surface, makes it possible to
observe the redistribution of the population around the Dirac cone by
elastic momentum scattering and to lower energies by inelastic scattering.
We have found that momentum scattering is much less efficient as compared to
inelastic scattering with strong supression of large-angle scattering,
as is expected from the chiral spin structure of the TSS.
V-doping has been shown to strongly enhance inelastic scattering, while
momentum scattering is almost unaffected, although the dopants carry a magnetic
moment. This is in agreement with
{\it ab initio} calculations which suggest that the scattering potential
of single magnetic impurities is dominated by its non-magnetic part.


\begin{thebibliography}{10}
\expandafter\ifx\csname url\endcsname\relax
  \def\url#1{\texttt{#1}}\fi
\expandafter\ifx\csname urlprefix\endcsname\relax\def\urlprefix{URL }\fi
\providecommand{\bibinfo}[2]{#2}
\providecommand{\eprint}[2][]{\url{#2}}

\bibitem{Xia09natphys}
\bibinfo{author}{Xia, Y.} \emph{et~al.}
\newblock \bibinfo{title}{{Observation of a large-gap topological-insulator
  class with a single Dirac cone on the surface}}.
\newblock \emph{\bibinfo{journal}{Nat. Phys.}} \textbf{\bibinfo{volume}{5}},
  \bibinfo{pages}{398--402} (\bibinfo{year}{2009}).

\bibitem{Chen09sci2}
\bibinfo{author}{Chen, Y.~L.} \emph{et~al.}
\newblock \bibinfo{title}{{Experimental realization of a three-dimensional
  topological insulator, Bi$_2$Te$_3$}}.
\newblock \emph{\bibinfo{journal}{Science}} \textbf{\bibinfo{volume}{325}},
  \bibinfo{pages}{178--81} (\bibinfo{year}{2009}).

\bibitem{Hasan10rmp}
\bibinfo{author}{Hasan, M.~Z.} \& \bibinfo{author}{Kane, C.~L.}
\newblock \bibinfo{title}{{Colloquium: Topological Insulators}}.
\newblock \emph{\bibinfo{journal}{Rev. Mod. Phys.}}
  \textbf{\bibinfo{volume}{82}}, \bibinfo{pages}{3045--67}
  (\bibinfo{year}{2010}).

\bibitem{Scanlon12am}
\bibinfo{author}{Scanlon, D.~O.} \emph{et~al.}
\newblock \bibinfo{title}{Controlling bulk conductivity in topological
  insulators: Key role of anti-site defects}.
\newblock \emph{\bibinfo{journal}{Advanced Materials}}
  \textbf{\bibinfo{volume}{24}}, \bibinfo{pages}{2154--2158}
  (\bibinfo{year}{2012}).

\bibitem{Wang13prb}
\bibinfo{author}{Wang, L.-L.} \emph{et~al.}
\newblock \bibinfo{title}{Native defects in tetradymite
  Bi${}_{2}$(Te${}_{x}$Se${}_{3\ensuremath{-}x}$) topological insulators}.
\newblock \emph{\bibinfo{journal}{Phys. Rev. B}} \textbf{\bibinfo{volume}{87}},
  \bibinfo{pages}{125303} (\bibinfo{year}{2013}).

\bibitem{Reimann18nat}
\bibinfo{author}{Reimann, J.} \emph{et~al.}
\newblock \bibinfo{title}{Subcycle observation of lightwave-driven dirac
  currents in a topological surface band}.
\newblock \emph{\bibinfo{journal}{Nature}} \textbf{\bibinfo{volume}{562}},
  \bibinfo{pages}{396} (\bibinfo{year}{2018}).

\bibitem{Kuroda16prl}
\bibinfo{author}{Kuroda, K.}, \bibinfo{author}{Reimann, J.},
  \bibinfo{author}{G{\"u}dde, J.} \& \bibinfo{author}{H{\"o}fer, U.}
\newblock \bibinfo{title}{Generation of transient photocurrents in the
  topological surface state of Sb$_2$Te$_3$ by direct optical excitation with
  mid-infrared pulses}.
\newblock \emph{\bibinfo{journal}{Phys. Rev. Lett.}}
  \textbf{\bibinfo{volume}{116}}, \bibinfo{pages}{076801}
  (\bibinfo{year}{2016}).

\bibitem{Russman19jpcs}
\bibinfo{author}{R{\"u}{\ss}mann, P.}, \bibinfo{author}{Mavropoulos, P.} \&
  \bibinfo{author}{Bl{\"u}gel, S.}
\newblock \bibinfo{title}{Lifetime and surface-to-bulk scattering off vacancies
  of the topological surface state in the three-dimensional strong topological
  insulators Bi$_2$Te$_3$ and Bi$_2$Se$_3$}.
\newblock \emph{\bibinfo{journal}{J. Phys. Chem. Solids}}
  \textbf{\bibinfo{volume}{128}}, \bibinfo{pages}{258 -- 264}
  (\bibinfo{year}{2019}).

\bibitem{Reimann14prb}
\bibinfo{author}{Reimann, J.}, \bibinfo{author}{G{\"u}dde, J.},
  \bibinfo{author}{Kuroda, K.}, \bibinfo{author}{Chulkov, E.~V.} \&
  \bibinfo{author}{H{\"o}fer, U.}
\newblock \bibinfo{title}{Spectroscopy and dynamics of unoccupied electronic
  states of the topological insulators Sb$_2$Te$_3$ and Sb$_2$Te$_2$S}.
\newblock \emph{\bibinfo{journal}{Phys. Rev. B}} \textbf{\bibinfo{volume}{90}},
  \bibinfo{pages}{081106} (\bibinfo{year}{2014}).
\newblock Erratum: \emph{Phys. Rev. B} \textbf{2015}, \emph{91} 039903.

\bibitem{Hajlaoui14natcomm}
\bibinfo{author}{Hajlaoui, M.} \emph{et~al.}
\newblock \bibinfo{title}{{Tuning a Schottky barrier in a photoexcited
  topological insulator with transient Dirac cone electron-hole asymmetry}}.
\newblock \emph{\bibinfo{journal}{Nat. Commun.}} \textbf{\bibinfo{volume}{5}},
  \bibinfo{pages}{3003} (\bibinfo{year}{2014}).

\bibitem{Niesner14prb}
\bibinfo{author}{Niesner, D.} \emph{et~al.}
\newblock \bibinfo{title}{{Bulk and surface electron dynamics in a p-type
  topological insulator SnSb$_2$Te$_4$}}.
\newblock \emph{\bibinfo{journal}{Phys. Rev. B}} \textbf{\bibinfo{volume}{89}},
  \bibinfo{pages}{081404} (\bibinfo{year}{2014}).

\bibitem{Sumida17scirep}
\bibinfo{author}{Sumida, K.} \emph{et~al.}
\newblock \bibinfo{title}{Prolonged duration of nonequilibrated dirac fermions
  in neutral topological insulators}.
\newblock \emph{\bibinfo{journal}{Sci. Rep.}} \textbf{\bibinfo{volume}{7}},
  \bibinfo{pages}{14080} (\bibinfo{year}{2017}).

\bibitem{Neupane15prl}
\bibinfo{author}{Neupane, M.} \emph{et~al.}
\newblock \bibinfo{title}{{Gigantic surface lifetime of an intrinsic
  topological insulator}}.
\newblock \emph{\bibinfo{journal}{Phys. Rev. Lett.}}
  \textbf{\bibinfo{volume}{115}}, \bibinfo{pages}{116801}
  (\bibinfo{year}{2015}).

\bibitem{Sumida19njp}
\bibinfo{author}{Sumida, K.} \emph{et~al.}
\newblock \bibinfo{title}{Magnetic-impurity-induced modifications to ultrafast
  carrier dynamics in the ferromagnetic topological insulators
  Sb$_{2\ensuremath{-}x}$V$_x$Te$_3$}.
\newblock \emph{\bibinfo{journal}{New J. Phys.}} \textbf{\bibinfo{volume}{21}},
  \bibinfo{pages}{093006} (\bibinfo{year}{2019}).

\bibitem{Sumida21pss}
\bibinfo{author}{Sumida, K.} \emph{et~al.}
\newblock \bibinfo{title}{Ultrafast surface dirac fermion dynamics of
  Sb$_2$Te$_3$-based topological insulators}.
\newblock \emph{\bibinfo{journal}{Prog. Surf. Sci.}}
  \textbf{\bibinfo{volume}{96}}, \bibinfo{pages}{100628}
  (\bibinfo{year}{2021}).

\bibitem{Kokh14}
\bibinfo{author}{Kokh, K.~A.}, \bibinfo{author}{Makarenko, S.~V.},
  \bibinfo{author}{Golyashov, V.~A.}, \bibinfo{author}{Shegai, O.~A.} \&
  \bibinfo{author}{Tereshchenko, O.~E.}
\newblock \bibinfo{title}{{Melt growth of bulk Bi$_2$Te$_3$ crystals with a natural
  p-n junction}}.
\newblock \emph{\bibinfo{journal}{Cryst. Eng. Comm.}}
  \textbf{\bibinfo{volume}{16}}, \bibinfo{pages}{581--84}
  (\bibinfo{year}{2014}).

\bibitem{Sobota12prl}
\bibinfo{author}{Sobota, J.~A.} \emph{et~al.}
\newblock \bibinfo{title}{{Ultrafast optical excitation of a persistent
  surface-state population in the topological insulator Bi$_2$Se$_3$}}.
\newblock \emph{\bibinfo{journal}{Phys. Rev. Lett.}}
  \textbf{\bibinfo{volume}{108}}, \bibinfo{pages}{117403}
  (\bibinfo{year}{2012}).

\bibitem{Gudde21pss}
\bibinfo{author}{G{\"u}dde, J.} \& \bibinfo{author}{H{\"o}fer, U.}
\newblock \bibinfo{title}{Ultrafast dynamics of photocurrents in surface states
  of three-dimensional topological insulators}.
\newblock \emph{\bibinfo{journal}{Prog. Surf. Sci.}}
  \textbf{\bibinfo{volume}{258}}, \bibinfo{pages}{2000521}
  (\bibinfo{year}{2020}).

\bibitem{Menshc11jetpl}
\bibinfo{author}{Menshchikova, T.~V.}, \bibinfo{author}{Eremeev, S.~V.} \&
  \bibinfo{author}{Chulkov, E.~V.}
\newblock \bibinfo{title}{{On the origin of two-Dimensional electron gas states
  at the surface of topological insulators}}.
\newblock \emph{\bibinfo{journal}{JETP Lett.}} \textbf{\bibinfo{volume}{94}},
  \bibinfo{pages}{106--11} (\bibinfo{year}{2011}).

\bibitem{Moser17jelsp}
\bibinfo{author}{Moser, S.}
\newblock \bibinfo{title}{{An experimentalist's guide to the matrix element in
  angle resolved photoemission}}.
\newblock \emph{\bibinfo{journal}{J. Electron Spectrosc.}}
  \textbf{\bibinfo{volume}{214}}, \bibinfo{pages}{29--52}
  (\bibinfo{year}{2017}).

\bibitem{Glinka15prb}
\bibinfo{author}{Glinka, Y.~D.}, \bibinfo{author}{Babakiray, S.},
  \bibinfo{author}{Johnson, T.~A.}, \bibinfo{author}{Holcomb, M.~B.} \&
  \bibinfo{author}{Lederman, D.}
\newblock \bibinfo{title}{Resonance-type thickness dependence of optical
  second-harmonic generation in thin films of the topological insulator
  $\mathrm{B}{\mathrm{i}}_{2}\mathrm{S}{\mathrm{e}}_{3}$}.
\newblock \emph{\bibinfo{journal}{Phys. Rev. B}} \textbf{\bibinfo{volume}{91}},
  \bibinfo{pages}{195307} (\bibinfo{year}{2015}).

\bibitem{Ketterl18prb}
\bibinfo{author}{Ketterl, A.~S.} \emph{et~al.}
\newblock \bibinfo{title}{{Origin of spin-polarized photocurrents in the
  topological surface states of Bi$_2$Se$_3$}}.
\newblock \emph{\bibinfo{journal}{Phys. Rev. B}} \textbf{\bibinfo{volume}{98}},
  \bibinfo{pages}{155406} (\bibinfo{year}{2018}).

\bibitem{Kuroda17spie}
\bibinfo{author}{Kuroda, K.}, \bibinfo{author}{Reimann, J.},
  \bibinfo{author}{G{\"u}dde, J.} \& \bibinfo{author}{H{\"o}fer, U.}
\newblock \bibinfo{title}{Momentum space view of the ultrafast dynamics of
  surface photocurrents on topological insulators}.
\newblock In \bibinfo{editor}{Betz, M.} \& \bibinfo{editor}{Elezzabi, A.~Y.}
  (eds.) \emph{\bibinfo{booktitle}{Ultrafast Phenomena and Nanophotonics XXI}},
  vol. \bibinfo{volume}{10102}, \bibinfo{pages}{101020Q--1}
  (\bibinfo{publisher}{SPIE Proceedings Series}, \bibinfo{address}{Washington},
  \bibinfo{year}{2017}).

\bibitem{Tanimura21prb}
\bibinfo{author}{Tanimura, H.}, \bibinfo{author}{Tanimura, K.} \&
  \bibinfo{author}{Kanasaki, J.}
\newblock \bibinfo{title}{Ultrafast relaxation of photoinjected nonthermal
  electrons in the \ensuremath{\Gamma} valley of GaAs studied by time- and
  angle-resolved photoemission spectroscopy}.
\newblock \emph{\bibinfo{journal}{Phys. Rev. B}}
  \textbf{\bibinfo{volume}{104}}, \bibinfo{pages}{245201}
  (\bibinfo{year}{2021}).

\bibitem{Dyck02prb}
\bibinfo{author}{Dyck, J.~S.}, \bibinfo{author}{Chen, W.},
  \bibinfo{author}{Uher, C.}, \bibinfo{author}{Drasar, C.} \&
  \bibinfo{author}{Lost'ak, P.}
\newblock \bibinfo{title}{{Heat transport in Sb$_{2\ensuremath{-}x}$V$_x$Te$_3$ single crystals}}.
\newblock \emph{\bibinfo{journal}{Phys. Rev. B}} \textbf{\bibinfo{volume}{66}},
  \bibinfo{pages}{125206} (\bibinfo{year}{2002}).

\bibitem{Sessi16natcom}
\bibinfo{author}{Sessi, P.} \emph{et~al.}
\newblock \bibinfo{title}{Dual nature of magnetic dopants and competing trends
  in topological insulators}.
\newblock \emph{\bibinfo{journal}{Nat. Comm.}} \textbf{\bibinfo{volume}{7}},
  \bibinfo{pages}{2041--1} (\bibinfo{year}{2016}).

\bibitem{Fauster07pss}
\bibinfo{author}{Fauster, T.}, \bibinfo{author}{Weinelt, M.} \&
  \bibinfo{author}{H{\"o}fer, U.}
\newblock \bibinfo{title}{Quasi-elastic scattering of electrons in
  image-potential states}.
\newblock \emph{\bibinfo{journal}{Prog. Surf. Sci.}}
  \textbf{\bibinfo{volume}{82}}, \bibinfo{pages}{224--243}
  (\bibinfo{year}{2007}).

\bibitem{Reuss99prl}
\bibinfo{author}{Reu{\ss}, C.} \emph{et~al.}
\newblock \bibinfo{title}{Control of the dephasing of image-potential states by
  {CO} adsorption on {Cu(100)}}.
\newblock \emph{\bibinfo{journal}{Phys. Rev. Lett.}}
  \textbf{\bibinfo{volume}{82}}, \bibinfo{pages}{153--156}
  (\bibinfo{year}{1999}).

\bibitem{Fauster00cp}
\bibinfo{author}{Fauster, T.}, \bibinfo{author}{Reuss, C.},
  \bibinfo{author}{Shumay, I.~L.} \& \bibinfo{author}{Weinelt, M.}
\newblock \bibinfo{title}{{Femtosecond two-photon photoemission studies of
  image-potential states}}.
\newblock \emph{\bibinfo{journal}{Chem. Phys.}} \textbf{\bibinfo{volume}{251}},
  \bibinfo{pages}{111--21} (\bibinfo{year}{2000}).

\bibitem{Russmann18diss}
\bibinfo{author}{Rüssmann, P.}
\newblock \emph{\bibinfo{title}{{S}pin scattering of topologically protected
  electrons at defects}}, vol. \bibinfo{volume}{173} of
  \emph{\bibinfo{series}{Schriften des Forschungszentrums Jülich Reihe
  Schlüsseltechnologien / Key Technologies}}
  (\bibinfo{publisher}{Forschungszentrum Jülich GmbH Zentralbibliothek,
  Verlag}, \bibinfo{address}{Jülich}, \bibinfo{year}{2018}).
\newblock \urlprefix\url{https://juser.fz-juelich.de/record/850306}.
\newblock \bibinfo{note}{Dissertation, RWTH Aachen University, 2018}.

\end{thebibliography}

\section*{Acknowledgements}
We gratefully acknowledge funding by the Deutsche
Forschungsgemeinschaft (DFG, German Research Foundation) through grant
numbers GU 495/2 and project ID 223848855-SFB~1083.
This work was further financially supported by KAKENHI (Nos. 17H06138 and 18H03683) and partly
supported by the bilateral collaboration program between RFBR (Russia,
No. 15–52-50017) and JSPS (Japan).
K.S. was financially supported by Grant-in-Aid for JSPS Fellows (Nos.
16J03874 and 19J00858).
O.E.T. and K.A.K. have been supported by the RFBR and DFG (project
No.21-52-12024) and by the Russian Science
Foundation (project No. 17-12-01047).

\section*{Author contributions statement}

J.G., A.K., and U.H. conceived the experiment, J.R., K.S., M.K., and J.G. carried out the experiment,
K.A.K. and O.E.T. grew the crystals and characterized their properties.
All authors analyzed the data and discussed the results.
J.G. wrote the manuscript with contributions from all authors.
All authors reviewed the manuscript.

\section*{Additional information}


\subsection*{Competing interests}
The authors declare no competing interests.


\end{document}